\documentclass[a4paper,prl,twocolumn,showpacs,10pt,aps,reprint]{revtex4-1}
\usepackage{amsmath,amssymb,hyperref}
\usepackage{graphicx}
\usepackage{datetime}
\usepackage{braket}
\usepackage[dvipsnames]{xcolor}
\usepackage[]{pdfpages}

\newcommand{\flo}[1]{{#1}}
\newcommand{\beencorrected}[1]{{#1}}
\newcommand{\figref}[1]{Fig.~\ref{#1}}

\newcommand{\micron}{\ensuremath{\mu{\rm m}}}
\newcommand{\gOne}{\ensuremath{g^{(1)}}}

\newcommand{\rrtt}{\ensuremath{{\bf r}, {\bf r'}, \tau }}

\newcommand{\supplement}{Supplementary Material}

\begin{document}

\title{Spatiotemporal coherence of non-equilibrium multimode photon condensates}

\author{Jakov Marelic}
\author{Lydia F. Zajiczek}\altaffiliation[Present address: ]{Analytical Science Division, National Physical Laboratory, Hampton Road, Teddington, Middlesex TW11 0LW, UK}
\author{Henry J. Hesten}
\author{Kon H. Leung}
\author{Edward Y. X. Ong}
\author{Florian Mintert}
\author{Robert A. Nyman}\email[Correspondence to
]{r.nyman@imperial.ac.uk} \affiliation{Physics Department,
Blackett Laboratory, Imperial College London, Prince Consort Road, SW7~2AZ, United Kingdom}

\date{\today}

\begin{abstract}
\flo{We report on the observation of quantum coherence of Bose-Einstein condensed photons in an \beencorrected{optically-pumped}, dye-filled microcavity. We find that coherence is long-range in space and time above condensation threshold, but short-range below threshold\beencorrected{, compatible with thermal-equilibrium theory.} Far above threshold, the condensate is no longer at \beencorrected{thermal equilibrium} and is fragmented over non-degenerate, spatially overlapping modes. A microscopic theory including cavity loss, molecular structure and relaxation shows that this multimode condensation is similar to multimode lasing induced by imperfect gain clamping.}
\end{abstract}

\maketitle


Quantum condensation and coherence are intimately linked for ensembles of identical particles.
Condensation, defined by a macroscopically large fraction of all particles being in a single state
(usually the ground state~\cite{Penrose56, PitaevskiiStringari}) is typically associated with coherence as seen in the first-order correlation function, which is proportional to the visibility of fringes of an interference measurement~\cite{MandelWolf}.

While observation of thermal equilibrium and macroscopic occupancy of the ground state are sometimes considered sufficient proof of Bose-Einstein condensation (BEC), the enhancement of coherence brought by BEC means that interferometry is one of the most urgent measurements to be made with a condensate~\cite{Andrews97, Bloch00}. Where thermal equilibrium is not completely reached, coherence is the defining characteristic of non-BEC quantum condensation, e.g for semiconductor exciton-polaritons~\cite{Kasprzak06, Deng06, Deng07, Balili07} and organic polaritons~\cite{Daskalakis14, Plumhof14}. In non-ideal Bose gases, such as ultracold atoms, interactions tend to reduce but not destroy the coherence~\cite{Castin97, Fattori08, Gustavsson08}.

Photon condensates in dye-filled microcavities are weakly-interacting~\cite{Snoke13, Nyman14, Chiocchetta14, VanDerWurff14}, inhomogeneous~\cite{Marelic15, Keeling16}, dissipative Bose gases close to thermal equilibrium at room temperature~\cite{Klaers10a, Klaers10b,Schmitt15,Kirton13,Kirton15}. It is worth noting that the physical system has some similarities to a dye laser, with the decisive difference being that lasing is necessarily a non-equilibrium effect whereas photons can also undergo BEC in thermal equilibrium. Consequently \beencorrected{BEC} implies macroscopic occupation of the ground state independently of the pump properties, whereas a laser is characterized by a large occupation of exactly the mode that is \beencorrected{most strongly} pumped~\cite{Fischer12}.


\beencorrected{
Unique among physical realisations of BEC, in dye-microcavity photon BEC the particles thermalise only with a bath and not directly among themselves. This implies that the establishment of phase coherence in the condensation process is necessarily mediated via indirect interactions through by the dye, i.e. a system whose fast relaxation renders all mediated interaction \textit{incoherent}. Notably, coherence in photon condensates has not yet been systematically measured.
}


Condensates with macroscopic occupation of two or more states without phase relation are called fragmented~\cite{Mueller06}. Whereas strong, attractive interactions favour fragmentation, repulsive interactions stabilise a single condensate mode~\cite{Leggett01, PitaevskiiStringari, Nozieres95}. Nevertheless, fragmentation has been observed using ultracold atoms in multiple spin states~\cite{DeSarlo13}, or separated spatial modes~\cite{Chen08, Krinner15}. Fragmented, dissipative condensates with spatially separated states have been seen in polaritons in semiconductors~\cite{Richard05,CerdaMendez10}, and organic solids~\cite{Daskalakis15}. It has been proposed that for driven, dissipative bosonic systems, multimode condensation is a general non-equilibrium phenomenon~\cite{Vorberg13}, when driving happens faster than dissipation (such as loss, thermal equilibration or spatial re-distribution).

Below threshold pump power, $P_{th}$, the coherence time $T$ and length $L$ of the thermalised light are expected to be of order $h/k_B T_0 \simeq 0.15$~ps and $\lambda_{dB}=\sqrt{h \lambda_0 c \,/\, 2\pi k_B T_0 n_L^2}\simeq 1.5\,\micron$ where $\lambda_0\simeq 590$~nm is the wavelength of the lowest-energy cavity mode, $T_0=300$~K the temperature, $c$ the speed of light in free space and $n_L$ the refractive index of the solvent filling the cavity~\cite{Guarrera11, deLeeuw13, deLeeuw14b}.
The coherence time is predicted to be much greater above threshold than below~\cite{deLeeuw14b, Kirton15}, increasing further as the number of particles in the condensate increases, and the coherence length is expected to be at least as large as the whole \beencorrected{condensate~\cite{deLeeuw14a}. Multimode condensation} may occur \flo{and} its effect on coherence is not predicted~\cite{Keeling16}.

In this manuscript we present measurements of the coherence properties of thermalised photons with both time delays and position shifts between the two arms of an interferometer. We describe how the controls and outputs of our imaging interferometer correspond to the underlying first-order correlation function,  $\gOne(\rrtt)$, as a function of positions ${\bf r}$ and ${\bf r'}$ and time delay $\tau$. We characterise the coherence time and length of the photon condensate as a function of pump power. Below and just above threshold, the measurements are compatible with thermal equilibrium theory. Far above threshold, the condensate fragments into multiple, spatially-distinct but overlapping, non-degenerate modes \flo{ accompanied by a decrease of both spatial and temporal coherence.} We interpret multimode condensation as a non-equilibrium phenomenon similar to gain saturation in lasers.

\begin{figure}	\centering
	\includegraphics[width=0.8\columnwidth]{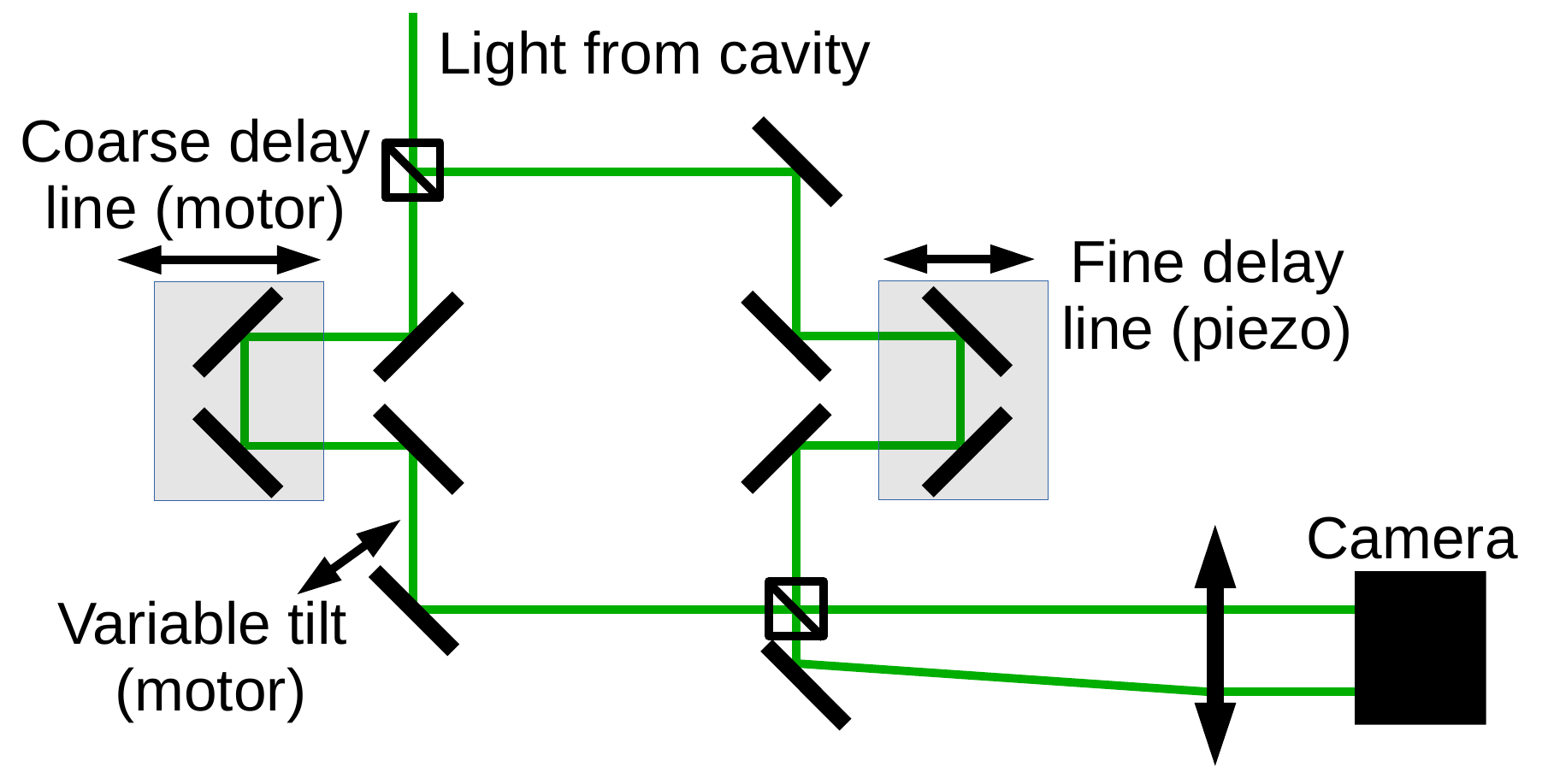}
	\\
	\includegraphics[width=0.98\columnwidth]{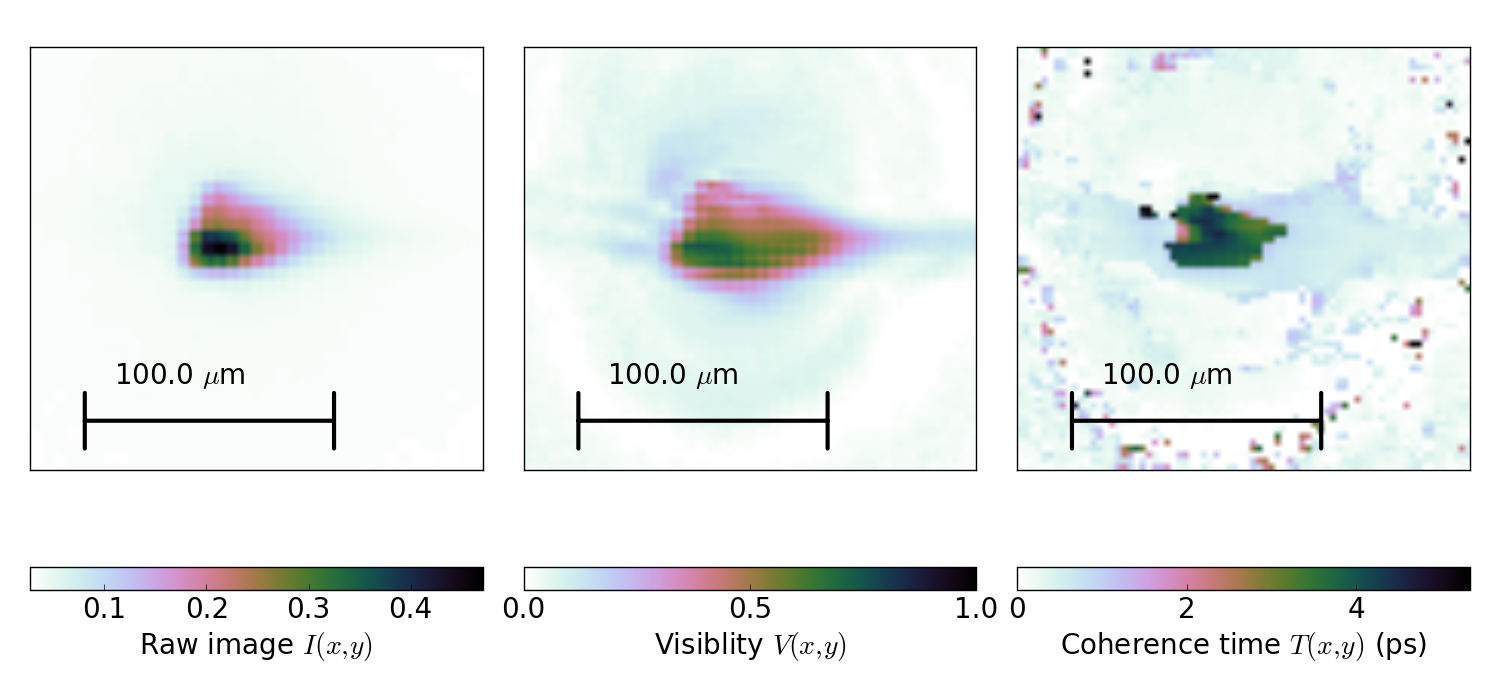}
	\caption{Top: Diagram of interferometer. The fine delay for scanning phase is controlled by a piezo actuator. Large-scale time delays $\tau$ are controlled by a motor. The image passing through one arm of the interferometer is shifted by a motorised mirror mount. Bottom: various levels of abstraction of the data just above threshold pump power, with overlapped images (${\bf r}={\bf r'}$). From left to right: a raw image at $\tau=0$, a visibility image (inferred from a set of 41 images at varying fine delay times), a coherence-time image (inferred from a set of 29 visibility images for varying $\tau$).}
	\label{fig:scheme}
\end{figure}

Our experiment starts by pumping a fluorescent dye in a high-finesse microcavity~\cite{Klaers10b, Marelic15} in quasi-continuous conditions. The pump spot was elliptical with a minor axis of typically 50--60~\micron\ diameter, and we use the 8\textsuperscript{th} longitudinal mode of the cavity with a cutoff wavelength of 590~nm. These parameters are known to produce near thermal-equilibrium conditions~\cite{Keeling16}. The cavity photoluminescence is imaged to infinity, then split. Half is split again and imaged onto an auxiliary camera and a spectrometer whose spectral resolution, about 0.2~nm, is insufficient to resolve the bare cavity modes which are separated by $0.05$~nm. The other half is sent to an imaging Mach-Zehnder interferometer, as shown in \figref{fig:scheme}. Each of the two arms of the interferometer has a delay line: one controlled by a piezo for the fine motion to scan over a fringe, the other controlled by a motor for coarse motion. The horizontal axis, $x$, of the last adjustable mirror in one arm is controlled by a motor, whose motion is converted to a shift in position of the image at the camera. Both outputs of the interferometer are sent onto a camera through a single imaging optic, imaged to two separate locations on the sensor. There is a linear-polarising filter in front of the camera, which increases the visibility of fringes. 

The camera records a spatially resolved intensity distribution. If one arm of the interferometer is blocked, this corresponds to the intensity $I({\bf r})$ emitted from the cavity, {\it i.e.} the spatial profile of the condensate photoluminescence. Since pumping and detection in this experiment are quasi-continuous, all processes are stationary. Temporal resolution comes in terms of the path delay of the interferometer. The detected interferometer signal depends on ${\bf r}$, ${\bf r'}$ and $\tau$, where \mbox{${\bf r}=(x,y)$} is the position on the camera, \mbox{${\bf r'}=(x+\delta_x,y)$}, with the displacement $\delta_x$ introduced by one arm of the interferometer and $\tau$ is the temporal delay corresponding to the path-length difference between the two arms of the interferometer. The Michelson visibility of fringes $V$ is directly related to the coherence \gOne:
\mbox{$
	 V(\rrtt) = 2\left|\gOne(\rrtt)\right| \sqrt{I({\bf r})I({\bf r'})} / \left[I({\bf r})+I({\bf r'})\right]
$}.

We scan the piezo-controlled delay, typically acquiring a set of 41 images, while maintaining all other parameters fixed. The principal result of our data analysis (explained in detail in \supplement) is a four-dimensional set of visibility data, $V(x,y,\delta_x, \tau)$.
\flo{For any value of $x$ and $y$ we can extract a characteristic coherence time $T$ (or length $L$), with a fit, usually to a Gaussian, in $\tau$ (or $\delta_x$ respectively).} In \figref{fig:scheme} (bottom, from left to right) we see an image of the photon condensate $I(x,y)$, an image of the visibility of the same condensate $V(x,y)_{\delta_x=0, \tau=0}$ and an image of its coherence time, $T(x,y)_{\delta_x=0}$. Since the images are overlapped ($\delta_x=0$), the visibility $V$ is exactly equal to the coherence \gOne.
\begin{figure}[hbt]
	\centering
	\includegraphics[width=0.49\columnwidth]{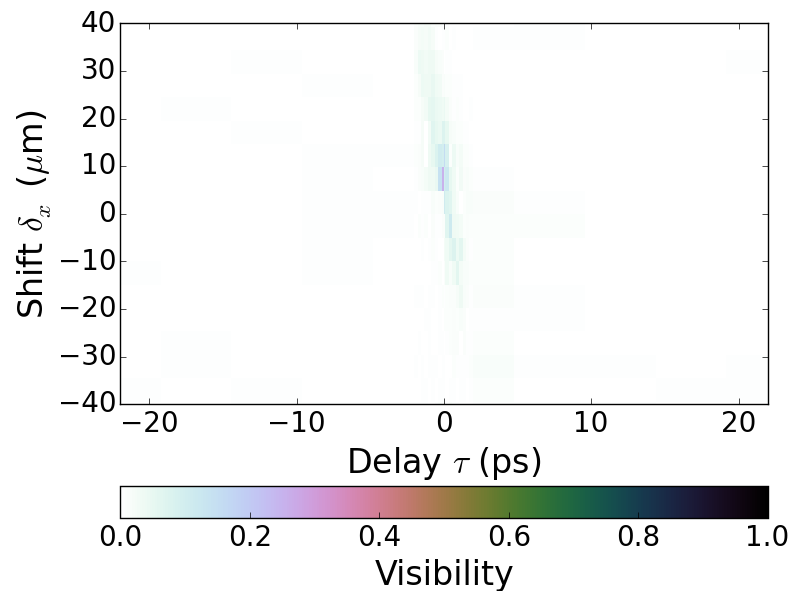}
	\includegraphics[width=0.49\columnwidth]{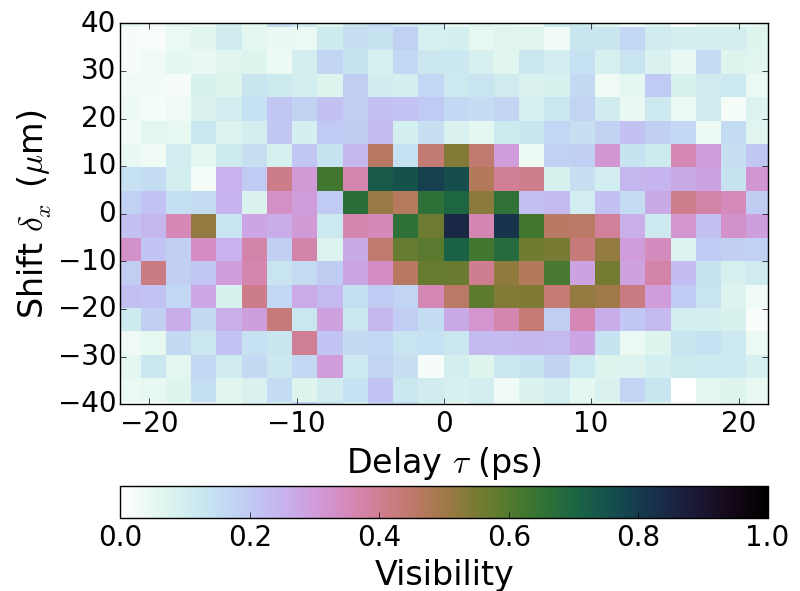}
	\caption{Visibility for a specific pixel, far below (left) and just above (right) threshold for condensation as delay and shift are varied, $V(\tau,\delta_x)_{x_0,y_0}$. The coherence times (lengths) are 0.2~ps (4.5~\micron) for the thermal cloud and 10~ps (14~\micron) for the condensate. }
	\label{fig:spatiotemporal}
\end{figure}
\figref{fig:spatiotemporal} is generated by choosing a single pixel $x_0,y_0$ and measuring the visibility $V(\tau,\delta_x)_{x_0,y_0}$  as a function of long-range delay and image shift. The results are shown just above and far below threshold. Under inspection, the differences between $V$ and \gOne\ were not noticeable, so we have presented $V$. \beencorrected{Coherence time and length are inferred from a two-dimensional Gaussian fit to the data.}

Far below threshold, the length and time scales of coherence, 4.5~\micron\ and 0.2~ps, are slightly longer than the thermal scales. This \beencorrected{overestimation} in both space and time is explained by finite spatial resolution of around 3~\micron\ (see \supplement). Above threshold the measured  length, 14~\micron, of the condensate is comparable to the size of the condensate itself, implying that the whole condensate shares one phase, as expected. The measured coherence time of 10~ps is also large, limited by condensate emission frequency fluctuations on timescales equal to the time between images, 200~ms. The condensate emission frequency variations are dominated by the variation of the cavity length at the limits of our locking scheme, which has a bandwidth of 20~Hz and resolution equivalent to about 0.05~nm in cavity cutoff wavelength~\footnote{These rapid cavity length fluctuations relate directly to the mechanical stability of the cavity. We optimise the mechanical stability through reinforcement screws, whose precise adjustment affects the maximum coherence time we observe from one experimental data set to another. We interpret all coherence times above 2~ps as ``large''.}.

\begin{figure}[htb]
	\centering
	\includegraphics[width=0.7\columnwidth]{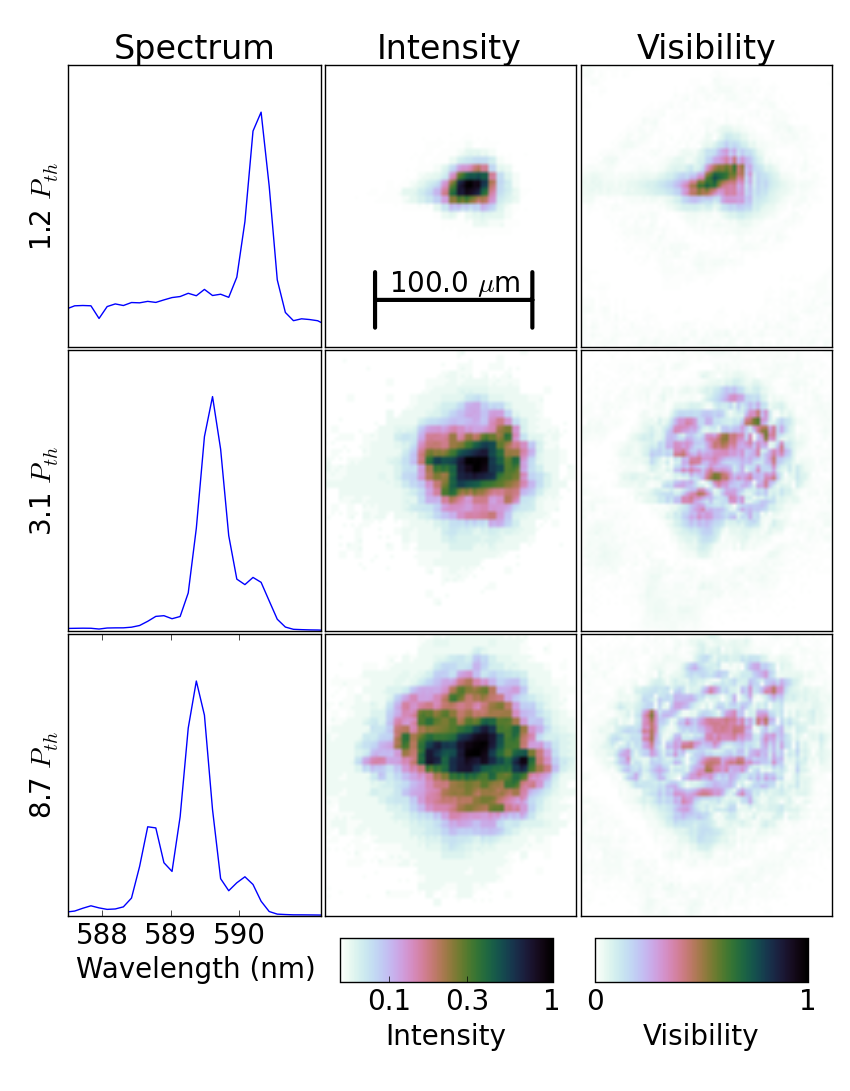}
	\caption{Normalised photoluminescence spectrum (left column), normalised image (middle column) and visibility image (right column) for various pump powers (rows, as labelled on the graph) above threshold $P_{th}$. The spectrum broadens and splits into multiple modes, the condensate broadens in space and the visibility image fragments at higher powers. A small pump spot ($30\pm 10$~\micron) was used to reduce threshold pump power.}
	\label{fig:multimode}
\end{figure}

\flo{ \figref{fig:multimode} depicts the spectrum, image $I({\bf r})$ and visibility image $V({\bf r})$ for various pump powers above threshold. The condensate peak in the spectrum broadens with increasing pump power and breaks up into multiple peaks, \beencorrected{i.e. the condensate splits into multiple non-degenerate modes. \figref{fig:multimode} shows three peaks, but we have seen up to five distinct peaks in some experimental runs where we used reduced pump spot sizes to lower the pump threshold.}}

Along with this non-degenerate multimode behaviour, the condensate broadens and the measured visibility \flo{develops a fragmented structure.}
\flo{\beencorrected{With a poorer spectrometer resolution, the spatial broadening} could have been taken as an indication of repulsive interactions \cite{Klaers10b}. Since, however, there is no blue-shift in the spectrum apart from variations of the cavity length~\cite{Note1}, \figref{fig:multimode} gives clear evidence of the condensation of non-interacting photons in several modes, \beencorrected{rather than quantum depletion reducing the condensate fraction}.}


We now ask: how does coherence vary across threshold and in the multimode regime? Are the non-degenerate modes coherent with each other, and is the multimode behaviour a sign of the breakdown of thermal equilibrium?

\begin{figure}[htb]
	\centering
	  \includegraphics[width=0.98\columnwidth]{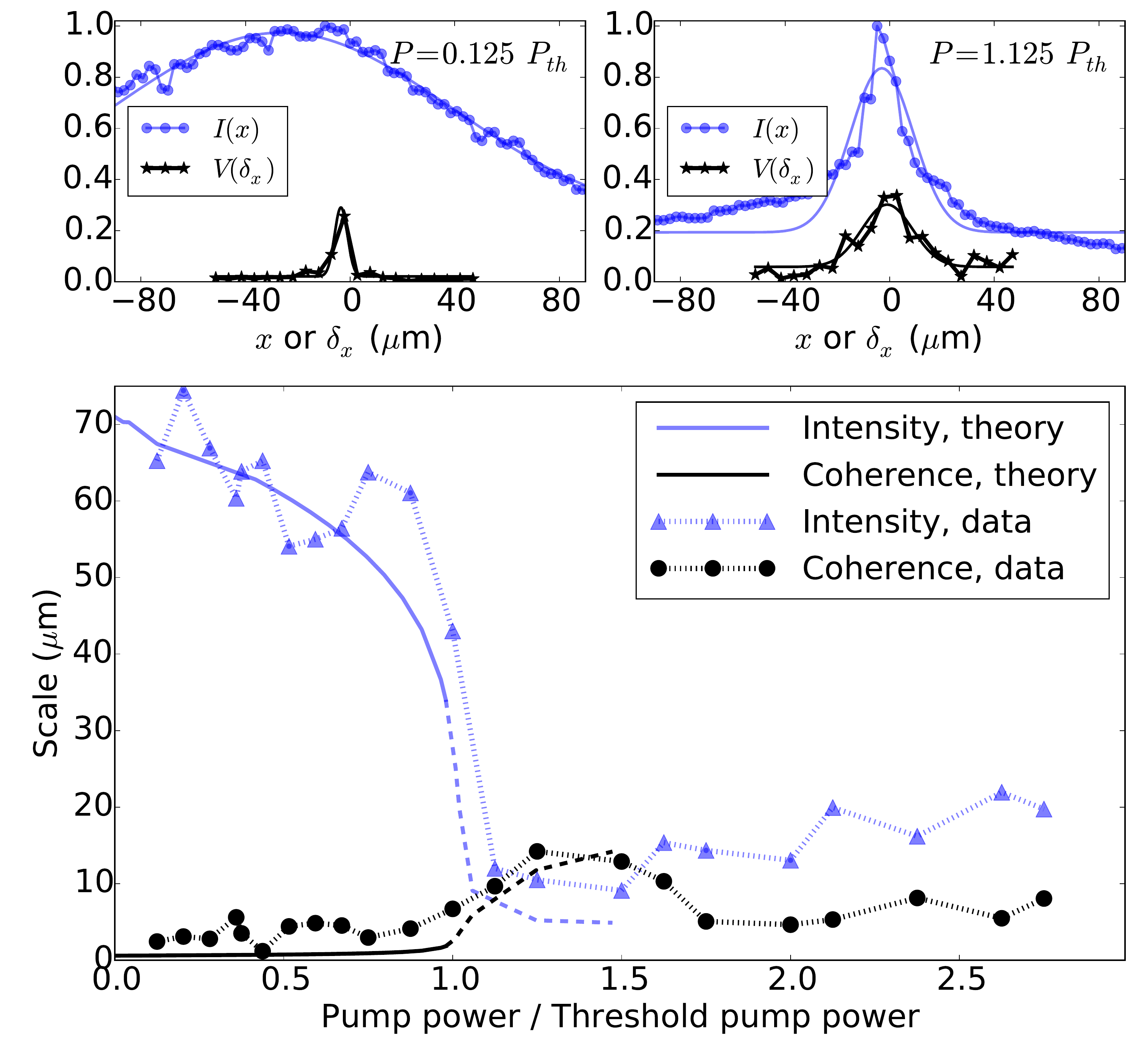}
	\includegraphics[width=0.98\columnwidth]{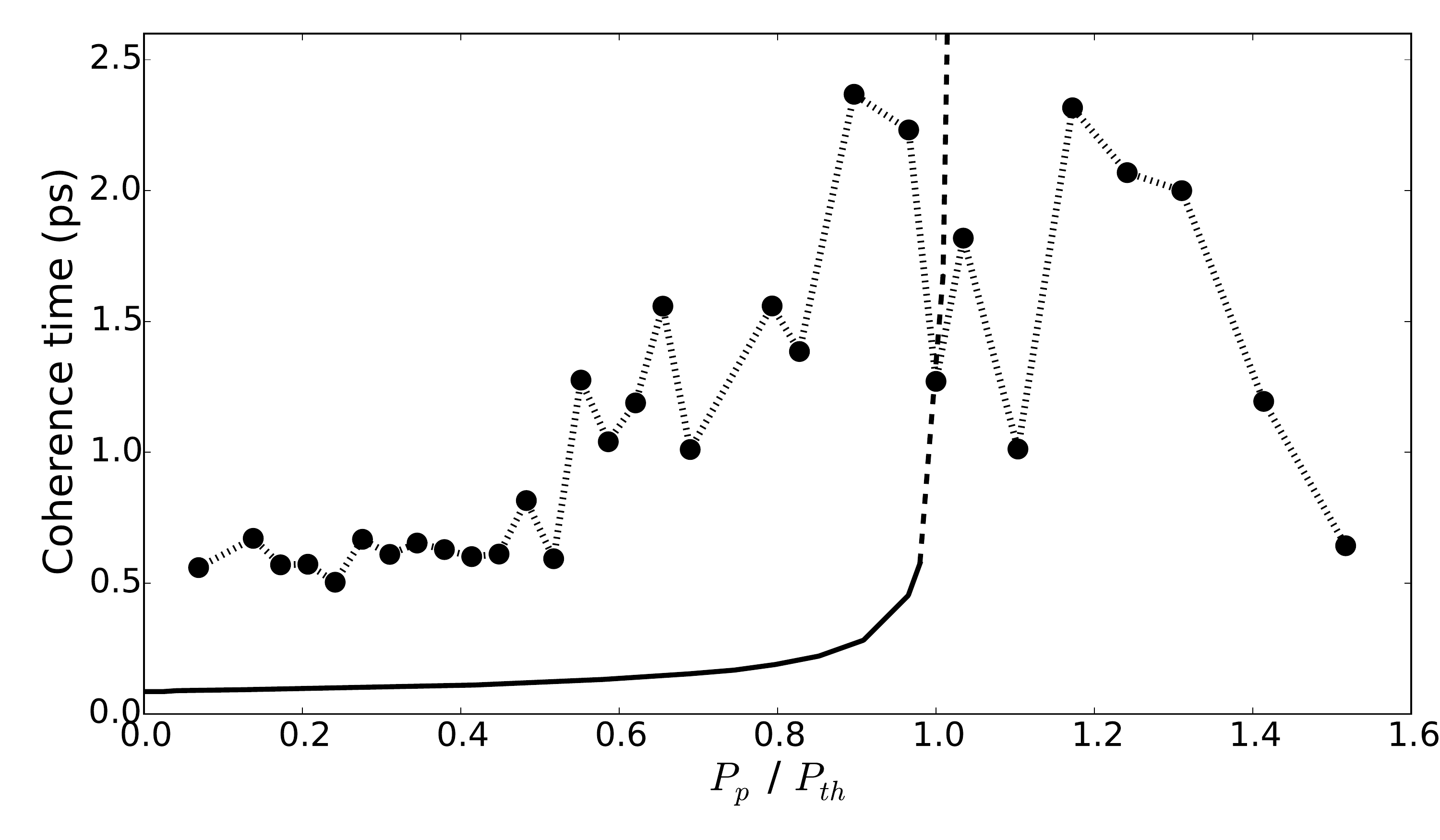}
	\caption{
	Coherence length and time as a function of pump power. Top: at pixel $(x_0, y_0)$, for two powers $P$, visibility as a function of shift between images $V(\delta_x,P)_{x_0,y_0,\tau=0}$  and intensity distribution $I(x,P)_{y_0}$ are shown, together with Gaussian fits. Middle: \beencorrected{size} of intensity distribution and visibility obtained from the fits. Solid lines are thermal-equilibrium theory with no adjustable parameters. Dashed lines are the same theory where it is only approximately valid. Bottom: coherence time, from Gaussian fits to $V(\tau, P)_{x_0,y_0,\delta_x=0}$. Solid line (dashed line) is the prediction assuming thermal equilibrium below threshold (and its approximate continuation above threshold).}
	\label{fig:multimode data}
\end{figure}

We define the spatial coherence length as the scale of a Gaussian fit to the visibility, as a function of shift between two images $V(\delta_x)$, and measure it for various pump powers. $V(\delta_x)$ and a cut through the photoluminescence intensity $I(x)$ are shown in \figref{fig:multimode data} (top), for two pump powers, one far below and one just above threshold. In \figref{fig:multimode data} (middle), we compare experiments to a thermal equilibrium theory without dissipation (see \supplement). The theory is based on a series expansion of the correlation function~\cite{Naraschewski99, Kohnen15} which agrees with exact calculations~\cite{Barnett00}. There are no free parameters below threshold (solid lines), but the scaling of the horizontal axis is imprecise above threshold (shown as dashed lines), as the number of photons varies non-linearly with pump power~\cite{Kirton15}.

\flo{Within the theory's range of validity ($P\lesssim P_{th}$) there is quantitative agreement with the experiment.}
Far below threshold, the coherence length is much shorter than the characteristic size of photoluminescence, limited only by imaging resolution. With increasing power around threshold, coherence length \flo{\beencorrected{grows, as the width of the emitted light decreases}. At even higher powers, when the system enters the multimode regime, the intensity increases but the coherence length decreases} to around 6~\micron\ (approximately the harmonic oscillator length scale), indicating that the multiple modes are incoherent. The condensate is only partially coherent, in contradiction to the dissipative, thermal-equilibrium prediction of Ref.~\cite{deLeeuw14a}.

\beencorrected{
In \figref{fig:multimode data} (bottom), we show the coherence time. Far below threshold, the shortest measured coherence time is limited by spatial resolution and marginal undersampling of the data~\footnote{The spatiotemporal coherence measurements of \figref{fig:spatiotemporal} below threshold do not suffer from undersampling and we observe as low as 0.24~ps coherence time. Spatial resolution still gives a lower limit.}. Above threshold, an upper bound for coherence time is set by the vibrations of the cavity~\cite{Note1}. Barely into the multimode regime, coherence time decreases, suggesting no coherence between modes, in agreement with the spatial coherence data. Even though thermal-equilibrium theory (black lines, solid below threshold, dashed above) does not describe the coherence time as accurately as spatial coherence and intensity, it captures qualitatively the increase of temporal coherence as the threshold pump power is reached.
%
}
 
\begin{figure}[htb]
	\centering
	\includegraphics[width=0.98\columnwidth]{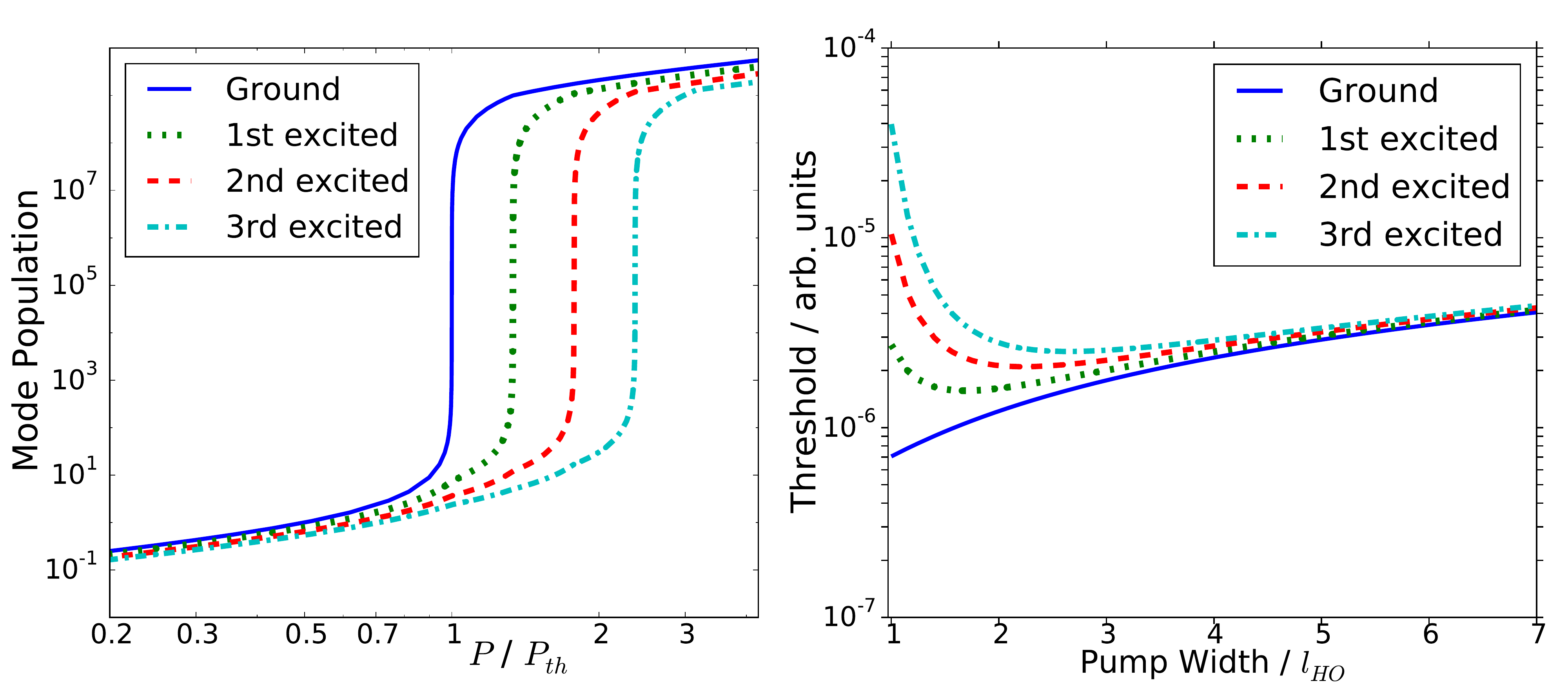}
	\caption{A microscopic model of dye molecules, cavity modes and dissipation explains the multimode behaviour. Left: for increasing pump power, first \beencorrected{the ground state reaches threshold, then more excited modes}. Right: threshold for different modes depends on the size of the pump spot, given in units of harmonic oscillator length $l_{HO}$.}
	\label{fig:multimode theory}
\end{figure}

\flo{Condensate fragmentation \beencorrected{cannot be explained by thermal-equilibrium processes.}
We therefore need to invoke a microscopic model that takes into account spatially-inhomogeneous pumping, molecular relaxation via the thermal bath of solvent vibrations, spontaneous emission and cavity loss \cite{Keeling16}.}
In \figref{fig:multimode theory} we show the results of the model (see \supplement\ for more details). For computation efficiency, the model is restricted to one dimension. With increasing pump power, condensation occurs  first in the lowest mode, then subsequently in higher modes (left panel). When one mode reaches threshold, it locally clamps the excited state population of dye molecules, but sufficient gain remains at the edges that more modes can reach threshold. The multimode regime is reached for the lowest pump powers for a pump spot which is large enough to overlap with several spatial modes of the bare resonator (right panel). \beencorrected{It is possible to extract approximate values for coherence length and time from the same microscopic model, and we find good qualitative agreement with \figref{fig:multimode data} in all regimes, below, near and far-above threshold.}


In conclusion, we have observed first-order coherence of thermalised photons in a dye-filled microcavity, below and above condensation threshold. Spatiotemporal correlations are longer-range for the condensed than non-condensed state, and show increases in range even below threshold, in broad agreement with thermal-equilibrium theory. \beencorrected{Above threshold, multiple modes are seen which is a signal of non-equilibrium, driven, dissipative processes~\cite{Vorberg13}.  There is no coherence between modes.  In this case, a microscopic model shows that the fragmentation can be explained using concepts from laser physics, i.e. imperfect gain clamping}.
By generating inhomogeneous, nonlinear gain and loss processes it may be possible to create equivalent states in other trapped condensates such as polaritons~\cite{Balili07, Winkler16} or atoms~\cite{Barontini13}. It would be intriguing to know if higher-order particle-particle correlations occur between modes even in the absence of phase coherence, and how superfluidity manifests itself in multimode condensates.

During review of this manuscript we became aware of related work on phase coherence of photon condensates~\cite{Schmitt16}. We thank Jonathan Keeling and Henk Stoof for inspiring discussions, and acknowledge financial support from the UK~EPSRC (via fellowship EP/J017027/1 and the Controlled Quantum Dynamics CDT EP/L016524/1), and the ERC (via ODYCQUENT grant).

\bibliographystyle{prsty}
\bibliography{photon_bec_refs}



\section*{Supplementary material}

# Spatiotemporal coherence of non-equilibrium multimode photon condensates: Supplementary material

Jakov Marelic, Lydia F. Zajiczek,* Henry J. Hesten, Kon H. Leung,

Edward Y. X. Ong, Florian Mintert, and Robert A. Nyman†

*Physics Department, Blackett Laboratory, Imperial College London,*
*Prince Consort Road, SW7 2AZ, United Kingdom*

(Dated: Thursday 18th February, 2016)

## S1. DATA ACQUISITION AND ANALYSIS

Our experimental apparatus is almost identical to [S1]. We pump a fluorescent dye in a high-finesse microcavity in quasi-continuous conditions, using 500 ns pulses, which are much longer than any thermalisation or cavity loss time scales in the system. The pulse repetition rate is varied so that the product of pump power and repetition rate is kept constant, up to 2000 mW laser output power, where the repetition rate is 500 Hz. Our maximum laser output power is 2200 mW, and about 50% of this light makes it into the cavity (through modulators and the cavity mirror). The repetition rate variation means that we have acceptable signal-to-noise over a very large range of pump powers. Images are integrated over, typically, 50–2000 ms. The pump spot was elliptical with an aspect ratio not far from one, and a minor axis between 50 and 60 $\mu$m diameter.

### A. Acquisition

The interferometer signal is observed using a colour camera. All colour values are converted to monochrome by summing red and green channels. For 590 nm (our typical working wavelength) the sensor of our camera (PointGrey Grasshopper GS3-U3-23S6C-C) is roughly equally sensitive in both red and green channels. Both output ports of the interferometer are directed to the same sensor. Two areas of the sensor are assigned as in-phase ($P$) and quadrature ($Q$). For each set of data, there is one image taken with one arm of the interferometer blocked, to allow us to align $P$ and $Q$ images.

To measure the visibility, we scan the voltage of the piezo controlling one of the delay lines while maintaining all other parameters fixed, so that the fine-scale delay time which we call $\tau_f$ varies over about 3 periods of oscillation of the light (6 fs). The piezo voltage is typically set to 41 values covering about 3 complete fringes. At each $\tau_f$ we take an image. Coarse-scale delays, $\tau$, up to 300 ps are achieved with the motor on the other delay line, with

---

* Present address: Analytical Science Division, National Physical Laboratory, Hampton Road, Teddington, Middlesex TW11 0LW, UK

† Correspondence to r.nyman@imperial.ac.uk

a precision of around 2 fs. The total delay is $\tau_f + \tau$, but we treat $\tau_f$ and $\tau$ as independent, since they vary over very different magnitudes, $|\tau_f| \ll |\tau|$. We overlap the images vertically (in the $y$ axis) but with a motorised mirror mount induce a shift in the horizontal axis, $\delta_x$. A full set of data for measuring the spatio-temporal coherence consists of images in $(x, y)$ for a variety of $\tau_f$, $\tau$ and $\delta_x$. Thus we build a five-dimensional data set $I(x, y, \delta_x, \tau, \tau_f)$. The notation includes the calibration of imaging magnification, so all co-ordinates presented are scaled to the intra-cavity co-ordinates.

A larger dataset is taken by varying control parameters of the photon condensate itself. In this manuscript, we only vary pump power $P_p$. Whenever the pump power is changed, the exposure and gain of the camera as well as the spectrometer are adjusted automatically to maximise dynamic range.

We make the reasonable assumption that the coherence varies only slowly compared to the oscillations of the light. The microscopic description of dye-microcavity photon condensates requires no processes faster than about 100 fs [S2–S6], which is much longer than the 2 fs oscillation period of the light, justifying this assumption.

### B. Analysis

When analysing data, each pair of $P$ and $Q$ images is aligned and binned if needed. Images are aligned by minimising the average-sum-square of differences between $P$ and $Q$ image values with respect to the shift of co-ordinates, taking only pixels which are present in both images after shifting. Knowing this optimised shift, we can be sure that a given pixel in $P$ corresponds to the same pixel in $Q$.

The intensities are $I_P(x, y, \delta_x, \tau, \tau_f)$ and $I_Q(x, y, \delta_x, \tau, \tau_f)$ for in-phase and quadrature respectively. In the analysis of data sets where the control parameters of the condensate (e.g. pump power $P_p$) vary, an array of values for each control parameter is constructed, e.g $I_P(x, y, \delta_x, \tau, \tau_f, P_p)$. The full six-dimensional data consists of as many 25 000 images, taking up to 30 GB of memory. Often 4-by-4 pixel blocks are combined to reduce computational effort in analysis. The major challenge is to visualise this data. This data set can be analysed and visualised in a number of ways, as shown in the main text, Fig. 1.

### 1. Visibility estimator

Extraction of the visibility over the set of fine delays is the most important processing we do on the data: $I_{P \text{ and } Q}(x, y, \delta_x, \tau, \tau_f, P_p) \rightarrow V(x, y, \delta_x, \tau, P_p)$. Our estimator for the visibility is based on a Fourier method and is robust against amplitude noise and converts phase and frequency noise to a reduction in visibility, unlike the Michelson visibility criterion. Amplitude noise is intrinsic to the photon condensate [S7]. Frequency noise is largely due to cavity length fluctuations. The result is, for each $P_p$, a four-dimensional visibility $V(x, y, \delta_x, \tau)$, which has been reduced over $\tau_f$.

Even averaging over much longer than cavity lifetime, there are some parameters of our experiment which are not well controlled, and so, close to threshold, there are large variations in photoluminescence intensity. Background noise (from readout of the sensor or background light) may also affect the interferometer outputs. The Michelson criterion of visibility (ratio of the differences between and the sums of maxima and minima of signal) is not robust against these kinds of noise. Phase noise is also present. To negate the effect of intensity fluctuations, we use arc-tangent of the ratio of the two quadratures:

$$\Phi(x, y, \delta_x, \tau, \tau_f) = \arctan\left[\frac{I_P(x, y, \delta_x, \tau, \tau_f)}{I_Q(x, y, \delta_x, \tau, \tau_f)}\right] \quad \text{(S1)}$$

In principle, it is important to subtract background signal and noise, but in practice we notice no effect of so doing.

The Michelson estimator for visibility is not robust because it does not make use of all the data available, only the maxima and minima. Instead we use of an estimator based on the Fourier transform. Using the autocorrelation of the arc-tangent data, we find the approximate frequency of the interference pattern. We then calculate the Fourier component of the arc-tangent signal at several frequencies more closely spaced than the Nyquist criterion, since the data typically only cover about 3 cycles. The Fourier component at angular frequency $\omega_f$ is:

$$\tilde{\Phi}(\omega_f) = \sum_j \Phi^{(j)} e^{i\omega_f \tau_f^{(j)}} \quad \text{(S2)}$$

for a sample of arctan data and fine delay times $\{\Phi^{(j)}, \tau_f^{(j)}\}$. The maximum amplitude gives the amplitude of fringes. The fringe amplitude is then normalised by half the sum of the data to give the visibility $V$, as would be expected for a clear sinusoid with an offset which gave Michelson visibility $V$.

We have tested the estimator using a model of noise in the system. We model the in-phase and quadrature signals, $I_P$ and $I_Q$ as

$$\tilde{I}_P = \left(A_0 + \tilde{A}\right)\left[\cos\left(\omega \tau_f + \tilde{\phi}\right)\right] + \tilde{R}_P + C \quad \text{(S3)}$$

$$\tilde{I}_Q = \left(A_0 + \tilde{A}\right)\left[\sin\left(\omega \tau_f + \tilde{\phi}\right)\right] + \tilde{R}_Q + C \quad \text{(S4)}$$

Variables with tildes are random variables. Phase noise $\tilde{\phi}$ and readout noises for each channel, $\tilde{R}_P$ and $\tilde{R}_Q$ are drawn from Gaussian distributions. We define an underlying amplitude $A_0$ and visibility $V_0$, which control the offset $C = A_0 V_0 / (1 - V_0)$. The angular frequency in piezo-controlled delay time units is $\omega$. The total amplitude is the sum of the underlying amplitude and amplitude noise, $\tilde{A}$. Amplitude noise in photon condensates is known to vary from normal to skewed with a long tail for large values, so we draw $\tilde{A}$ from a scaled Poisson distribution.

We tested four types of estimators against this noise model: the Michelson criterion, a root-mean-square test, sinusoid-fitting and the Fourier-based method described above. The Michelson criterion is not robust against noise, and the root-mean-square method requires scanning over an exact integer number of fringe periods. Least-squares fitting a sine wave gave fits were not robust, often picking local not global optima. Finally, the Fourier-based method was found to be robust (especially against amplitude noise when using the arc-tangent data) and low-noise, although we find that it is slightly biased. For extreme values of underlying visibility (near zero or unity), the estimator is biased towards 0.5. The extent of the bias depends on the amplitude of the noises. For experimentally realistic parameters, from an underlying visibility of 1.0 our estimator gives about 0.8. If the detector is weakly saturated, then the maximum inferred visibility may be further reduced. We do not correct for this in our presented data, as we are not fully certain that our noise model is complete, nor do we know the exact parameters that match our experiment. Also, the uncertainties in the inferred visibilities are about as large as the correction required to compensate for the bias.

## S2. UNCERTAINTIES

The data presented in the main body of the manuscript are presented without error bars. There are two major sources of uncertainty in our experiments: fluctuations of threshold, and of visibility. The fluctuations are slightly faster than the typical time to acquire a data set, 1–5 hours.

### A. Variability near threshold

In Fig. S1, we vary the pump power and measure the output light intensity averaged over a small region at the center of the image. The pump spot was smaller than use in the main text, to allow us to reach pump powers far above threshold here. Camera exposure and gain are set automatically for each power to avoid saturation. Pulse repetition rate was held constant at 500 Hz for this specific data set. Exposure is at least 2 ms, i.e. at least one 500 ns quasi-CW pump pulse is always detected. Below threshold, the signal to noise is less than unity. Above

threshold, the intensity grows linearly with pump power. A bi-linear fit reveals that the pump power at threshold is 260 mW. The measured threshold varies from one experimental run to another. However, many points below threshold show a condensate. In the range 180-250 mW, there may be a condensate or not. We do not know the cause of this variability. It is not the variation of cavity cutoff wavelength (although that does fluctuate on 10-minute time scales). It may be related to polymerisation of the dye, which we know occasionally forms clumps requiring cleaning of the cavity. Although the data might appear to show hysteresis (memory effects), the pump power was varied in a random order.

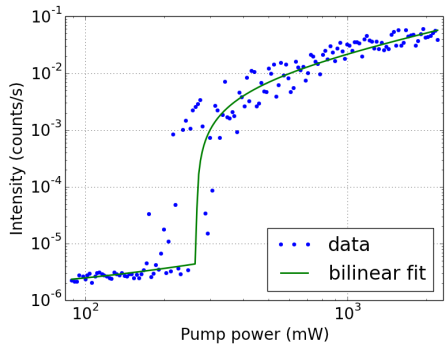

FIG. S1. Threshold behaviour. A bilinear fit shows threshold around 260 mW, but many points in the range 180–250 mW show condensation. This variability in threshold limits how reliably the experiment can operate close to threshold.

In the main manuscript, where power is varied, we reject data which are below threshold when they are expected to be above it, and vice versa. We accept only data that follow an approximately monotonic increase in output intensity as a function of pump power.

### B. Visibility variability

Standard-deviation errors are of limited use near threshold as there is no reason to believe that visibility measurements for a given set of parameters are drawn from a normal distribution. They are as likely to be drawn from a bimodal distribution, corresponding to below- and above-threshold behaviour. We have tried to ascertain the limiting uncertainties in visibility away from threshold, by measuring a large sample of visibilities as a function of delay above threshold: see Fig. S2.

By oversampling, we can build sub-samples and evaluate their standard deviations. The highest sub-sample-averaged visibility measured is about 0.7, although we know that our estimator is low-biased for such large visibilities. The largest shot-to-shot uncertainty in visibility is 0.15. For lower average visibility, the uncertainty in visibility is lower. For example, we can measure non-zero visibility of 0.04 with a signal-to-noise of unity in a single measurement. Our noise model produces a

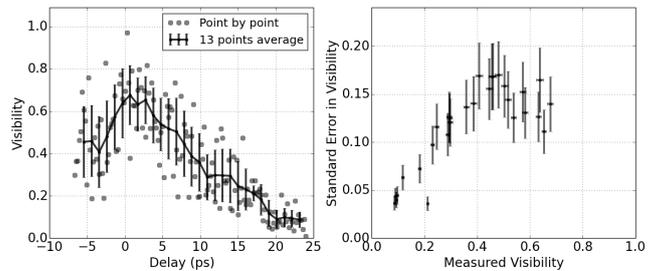

FIG. S2. Limiting uncertainty in the visibility, significantly above threshold. Left: Visibility variation with delay for overlapped images as a specific pixel. Grey dots are individual data points. Points with error bars are averages over 13 points centred on the marker, with error bar being the standard deviation of that sample. Right: Standard deviation of visibility as a function of mean visibility. Maximum visibility inferred is 0.7. Our estimator is low-biased for the largest visibilities, and high-biased for the smallest visibility values.

similar pattern, i.e. maximum inferred visibility 0.7 and standard deviation $> 0.1$, only with unrealistically large phase noise, of amplitude at least $\pi/4$ radians. We conclude that there is intrinsic noise in the visibility which does not come from our measurement apparatus or visibility estimator.

## S3. THERMAL EQUILIBRIUM THEORY OF BOSE GAS COHERENCE

The theory in the main text Fig. 4 is derived assuming a non-interacting Bose gas at thermal equilibrium in a symmetric, two-dimensional, harmonic trapping potential in the grand canonical ensemble [S8, S9]. We have also made use of an extension of the theory from spatial to temporal correlations [S10] which assumes that dissipative processes play no role. The theory is based on a series expansion in the fugacity. Fugacity is defined as $\zeta = \exp(\mu/k_B T)$, where $\mu$ is the chemical potential. The $k^{\text{th}}$ term in the expansion corresponds to occupancy of up to $k$ particles in any given mode. Far below threshold, very few terms are needed for the series to converge. Arbitrarily large phase-space density can be obtained for negative chemical potential, i.e. $\zeta < 1$, so the infinite series will always converge. The finite series also converges for large numbers of terms, albeit with a small positive chemical potential ($\zeta - 1 \ll 1$) for large particle numbers. First-order correlations, normalised or otherwise, are calculated using equations (20)–(23) in Ref. [S10].

Coherence length is defined here as the size of a Gaussian function which fits the visibility as a function of shift $V(\mathbf{r_0}, \mathbf{r_0} + \hat{\mathbf{x}}\delta_{\mathbf{x}}, \tau = 0)$, and not a fit to $|g^{(1)}(\mathbf{r_0}, \mathbf{r_0} + \hat{\mathbf{x}}\delta_{\mathbf{x}}, \tau = 0)|$ (the normalised first-order correlation function), which makes comparison to experimental data more robust against noise, especially in low-intensity regions of the images. Intensity scale is a Gaussian fit to the number density of photons as a

function of position. Number density can be extracted easily from the theory since number density at position $\mathbf{r}$ is equal to $G^{(1)}(\mathbf{r}, \mathbf{r}, \tau = 0)$ (the un-normalised correlation function). Coherence time is the full-width at half-maximum of $g^{(1)}(\mathbf{r_0}, \mathbf{r_0}, \tau)$. The correlation function in space matches well to a Gaussian, but the density does not. The temporal correlation function does not match any simple function (Gaussian, Lorentzian or symmetric double exponential decay), not least because, without damping, there are revivals of correlations functions at half-periods of the oscillations in the harmonic trapping potential. Real temporal coherence data is fitted with Gaussians.

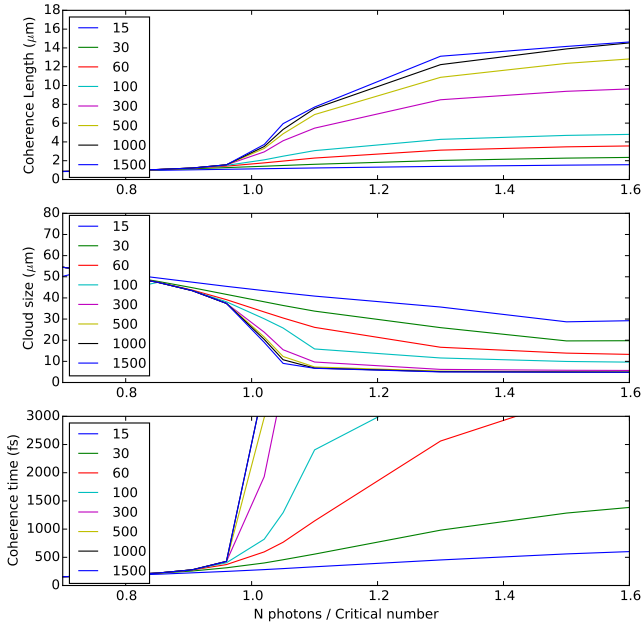

FIG. S3. Demonstration that the series expansion used for the theory in the main text converges even above threshold. The numbers in the legend refer to the number of terms used in the series. In the main text, Fig. 4, 999 terms are used.

In Fig. S3 we show how well the finite series expansion converges, for coherence length, intensity length scale and coherence time. Far below threshold, the series converges as expected. Above threshold, the spatial scales converge when the series has a number of terms of the same order as the condensate population. The results qualitatively agree with exact calculations [S9].

The temporal correlation functions converge only below threshold. There is qualitative agreement with the dissipative model of Kirton and Keeling [S2, S11], in that coherence time does increase slightly with increasing number (comparable to the pumping rate) as threshold is approached. Since the model used takes no account of dissipation, it cannot predict anything but that the coherence time of a pure condensate ought to be infinite. The maximum coherence time inferred here is $> 2.5$ ps, limited by the range over which $g^{(1)}$ is evaluated, avoiding the non-observed revivials.

While there are no adjustable parameters in the theory, the theoretical photon number does not directly correspond to the experimental pump power. Below threshold, photon number and pump power are experimentally seen and theoretically expected [S11] to be linearly proportional. Likewise, far above threshold, but not so just above threshold. We can therefore trust our calculations only below threshold, and above threshold our calculations (assuming that the photon number remains proportional to the pump power) are plotted as dashed lines and are only a qualitative guide to what is expected. The calculations presented in the main text, Fig. 4, use 999 terms of the expansion.

## A. Finite spatial resolution

The effect of finite imaging resolution and numerical aperture on measured interference patterns can be taken into account, starting from the known electric field at a point, $E(\mathbf{r})$. Finite resolution is imposed by convolving with a point-spread function $F(\mathbf{R})$:

$$\overline{E}(\mathbf{r}, t) = \int \mathrm{d}^2\mathbf{R} \, E(\mathbf{r} - \mathbf{R}, t) F(\mathbf{R}) \qquad (S5)$$

The overline indicates that finite resolution has been applied. The effect of finite numerical aperture is equivalent to applying the Fourier transform, applying a cutoff (multiplying by a top-hat function) and then inverse transforming, i.e. convolution with a cardinal sine, $\sin(x)/x$. This function can then simply be absorbed in the definition of the point-spread function, $F$.

The light at one output port of the interferometer is $\overline{E}_P(\mathbf{r}, \mathbf{r}', \tau) = \frac{1}{\sqrt{2}} \left[ \overline{E}(\mathbf{r}, t) + \overline{E}(\mathbf{r}', t') \right]$, where as usual $\tau = t - t'$. The effects of finite resolution are applied before the interference. The other output, $Q$, takes a minus instead of a plus. Then the intensity is:

$$2I_P(\mathbf{r}, \mathbf{r}', \tau) = \left\langle \left[ \overline{E}^\dagger(\mathbf{r}, t) + \overline{E}^\dagger(\mathbf{r}', t') \right] \left[ \overline{E}(\mathbf{r}, t) + \overline{E}(\mathbf{r}', t') \right] \right\rangle$$
$$= \langle \overline{E}^\dagger(\mathbf{r}, t) \overline{E}(\mathbf{r}, t) \rangle + \langle \overline{E}^\dagger(\mathbf{r}', t') \overline{E}(\mathbf{r}', t') \rangle$$
$$+ 2\mathrm{Re} \left[ \langle \overline{E}^\dagger(\mathbf{r}, t) \overline{E}(\mathbf{r}', t') \rangle \right] \qquad (S6)$$

The first two terms are the intensities as seen with finite resolution. The last term written more explicitly is:

$$\langle \overline{E}^\dagger(\mathbf{r}, t) \overline{E}(\mathbf{r}', t') \rangle = \overline{G^{(1)}}(\mathbf{r}, \mathbf{r}', \tau)$$
$$= \left\langle \int \mathrm{d}^2\mathbf{R} \, E^\dagger(\mathbf{r} - \mathbf{R}, t) F^*(\mathbf{R}) \int \mathrm{d}^2\mathbf{R}' E(\mathbf{r}' - \mathbf{R}', t') F(\mathbf{R}') \right\rangle$$
$$= \iint \mathrm{d}^2\mathbf{R} \mathrm{d}^2\mathbf{R}' F^*(\mathbf{R}) F(\mathbf{R}') \langle E^\dagger(\mathbf{r} - \mathbf{R}, t) E(\mathbf{r}' - \mathbf{R}', t') \rangle$$
$$= \iint \mathrm{d}^2\mathbf{R} \mathrm{d}^2\mathbf{R}' \, F^*(\mathbf{R}) F(\mathbf{R}') G^{(1)}(\mathbf{r} - \mathbf{R}, \mathbf{r}' - \mathbf{R}', t - t')$$
$$(S7)$$

It is possible to calculate the equilibrium first-order correlation function $G^{(1)}(\mathbf{r}, \mathbf{r}', \tau)$ for a non-condensed

Bose gas in a harmonic trap if we make the strong approximations that there are no dissipative processes and that thermal equilibrium is respected [S10]. Using the symbols defined in Ref. [S10]: $s$ is a co-ordinate $x$ or $y$, $K_s^{(k)}(s, s', t, t')$ is the propagator, $\zeta$ the fugacity and $\beta = 1/k_B T$ the inverse temperature we obtain:

$$\overline{G^{(1)}}(\mathbf{r}, \mathbf{r}', \tau) = \tag{S8}$$
$$\sum_{k=1}^{\infty} \zeta^k \prod_{s=x,y} \iint \mathrm{d}S\, \mathrm{d}S'\, F_s^*(S) F_s(S') \times$$
$$K_s^{(k)}\left(s - S, s' - S', t, [t' - \mathrm{i}k\hbar\beta]\right)$$

where we have also assumed that point-spread function is separable: $F(\mathbf{R}) = F_x(X) F_y(Y)$. This expression can be evaluated numerically, either by direct integration or via Fourier transforms. The finite-resolution correlation function is normalised:

$$\overline{g^{(1)}}(\mathbf{r}, \mathbf{r}', \tau) = \frac{\overline{G^{(1)}}(\mathbf{r}, \mathbf{r}', \tau)}{\sqrt{\overline{G^{(1)}}(\mathbf{r}, \mathbf{r}, 0)\, \overline{G^{(1)}}(\mathbf{r}', \mathbf{r}', 0)}} \tag{S9}$$

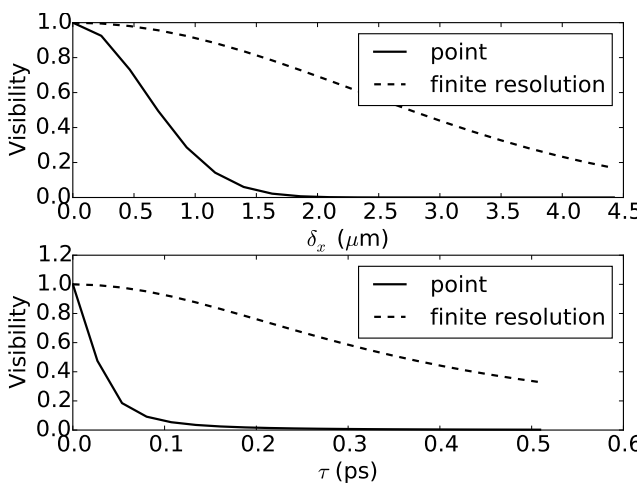

FIG. S4. Calculated effect of finite imaging resolution on $g^{(1)}$ for the non-condensed photons, equivalent to $P \simeq 5 \times 10^{-3} P_{th}$, using a 3 $\mu$m imaging resolution. The series expansion used 5 terms, which was sufficient for convergence so far below threshold.

The results are shown in Fig. S4 for a thermal cloud with pump power far below threshold ($P \simeq 5 \times 10^{-3} P_{th}$). The point correlation functions show shorter range coherence than those integrated over a finite resolution (a rotationally symmetric, 3 $\mu$m Gaussian point-spread function), and are consistent with the results seen at low pump powers in the main text, Figs. 2, 4 and 5.

## S4. NON-EQUILIBRIUM THEORY OF PHOTON THERMALISATION AND CONDENSATION

Since the thermal-equilibrium theory clearly breaks down in the multimode regime, we have implemented the non-equilibrium model of Kirton and Keeling [S2, S11, S12]. To explain the multimode behaviour we are obliged to treat the spatial dependence of pumping. In Ref. [S12] it is suggested that the multimode behaviour might occur with inhomogeneous pumping, because of imperfect clamping of the excited-state fraction of dye molecules. To study the threshold for various modes, we evaluate Ref. [S12] equations (8) and (9). A crucial quantity in those equations is the absolute value of light-matter coupling, denoted by $\Gamma$.

The light-molecule coupling strength can be approximately derived from measured quantities, noting that the typical timescale for population variations is:

$$\tau_{typ} = N\Gamma_{0D}(\lambda) = \rho_{1D}\Gamma_{1D}(\lambda) = \rho_{2D}\Gamma_{2D}(\lambda) \tag{S10}$$

where $N$ is the number of molecules, $\Gamma_{dD}$ and $\rho_{dD}$ are the light-matter coupling and molecular density in $d$ dimensions and $\lambda$ the wavelength of light in vacuum. In three dimensions, the mean time between scattering events for photons moving at speed $c^*$ from molecules at volume density $\rho_{3D}$ is $\tau_{typ} = 1/\rho_{3D} c^* \sigma(\lambda)$. Here $\sigma(\lambda)$ is the scattering cross-section at wavelength $\lambda$. Equating the timescales we find $\Gamma_{3D}^{max} = \sigma(534 \text{ nm}) c^* = 5.1 \times 10^{-12} \text{ m}^3/\text{s}$. The variation of light-matter coupling with wavelength is known by interpolating experimental measurements enforcing a Kennard-Stepanov relations [S12]. In lower dimensions, the density is scaled typically by the cavity physical length and/or the harmonic oscillator length.

### A. Multimode behaviour

The principal results are shown in the main text, Fig. 5, for the following parameters (symbols as used in Ref. [S12]) in one dimension:

- Pump spot size (left panel of main text Fig. 5): 2.2 $l_{HO}$, which is slightly smaller than the experiment (approx 10 $l_{HO}$), which compensates for the fact that we cannot efficiently perform the computations in two dimensions.

- Mode spacing: $\hbar \times 3 \times 10^{14}$ rad s$^{-1}$ equal to that used in the experiment.

- Number of modes: 15, sufficient to make the main results converge.

- Cavity cutoff (lowest mode detuning from resonance): $3 \times 10^{14}$ rad s$^{-1}$ which is equivalent to 596 nm.

- Cavity decay rate: $\kappa = 10^9$ s$^{-1}$

- Molecular density: $\rho_{1D} = 5.1 \times 10^{12}$ molecules/$l_{HO}$ which is equivalent to 1.7 mM solution concentration (similar to the experiments) if the extra dimension for conversion from 2D to 1D is taken to be the harmonic oscillator length, 6 $\mu$m. The length for conversion from 3D to 2D is the physical space between the mirrors, $(q - q_0)\lambda^*/2$ with $q = 8$ being the longitudinal mode number and $q_0 \simeq 4$ expresses how far the electric field penetrates the surface of the dielectric mirrors.

- Spontaneous emission lifetime: $\Gamma_\downarrow = 0.25 \times 10^9 \text{ s}^{-1}$ which is the known value for Rhodamine 6G in ethylene glycol.

## B. Coherence length and time

The linewidth theory of Ref. [S11] takes into account only the single mode into which condensation occurs. Such a truncation is not appropriate unless mode filtering optics are used. Instead, we make the approximation that dissipative processes are negligible, which means that we can write the classical electric field of the light in the cavity as a coherent sum over electric fields from the many modes:

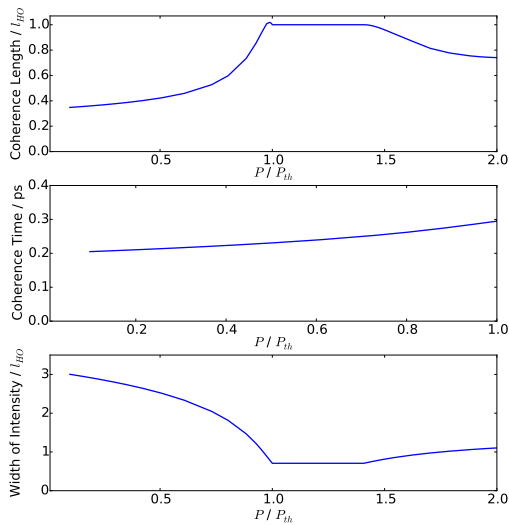

FIG. S5. Inhomogeneous model coherence length and time from microscopic theory. These results come from Eqn. (S13) and S13 and reproduce qualitatively the experimental results of the main text, Fig. 4. The parameters here are the same as those used in the main text Fig. 5, except for using 45 modes here.

$$E(\mathbf{r}, t) = \sum_m \psi_m(\mathbf{r}) e^{-i\omega_m t} \sqrt{n_m} \quad (S11)$$

where the population of mode $m$ is $n_m$, its associated eigenfunction and angular frequency are $\psi_m(\mathbf{r})$ and $\omega_m$. The interferometer detectors can be used to measure the first order correlation function, much as in Eqn. (S6), but remembering that only stationary processes are observed. The result is:

$$G^{(1)}(\mathbf{r}, \mathbf{r}', \tau) = \lim_{T \to \infty} \frac{1}{T} \int_0^T dt \, E^*(\mathbf{r}, t) E(\mathbf{r}, \mathbf{r}', t, t+\tau) \quad (S12)$$

Note that since the electric field we wrote was classical, there is no quantum averaging here.

Simple formulae that follow are:

$$G^{(1)}(\mathbf{0}, \mathbf{0}, \tau) = \sum_m |\psi_m(\mathbf{0})|^2 \, \text{Re}\left[e^{-i\omega_m \tau}\right] n_m \quad (S13)$$

$$G^{(1)}(\mathbf{0}, \mathbf{r}', 0) = \sum_m \text{Re}\left[\psi_m^*(\mathbf{0})\psi_m(\mathbf{r}')\right] n_m \quad (S14)$$

Results from these formulae are plotted in Fig. S5, summarised as the experimental data is, using Gaussian fits to extract coherence lengths and times, as well as real-space extents, $G^{(1)}(\mathbf{r}, \mathbf{r}, 0)$. For numerical reasons we cannot evaluate with enough modes to compare quantiatively to the experimental data, but the qualitative agreement is very good. For increasing pump powers, the spatial extent falls until threshold is reached, then rises in the multimode regime. Likewise the spatial coherence length rises then falls. The temporal coherence grows but above single-mode condensation threshold our approximation ignoring dissipative processes is not valid, until the multimode regime is reached. Agreement with the thermal-equilibrium model below threshold is good, and there is qualitative agreement with experimental results in all regimes (below, near and far-above threshold).



\end{document}